\documentclass[a4paper,prl, superscriptaddress, twocolumn,longbibliography]{revtex4-2}

\usepackage[utf8]{inputenc}
\usepackage[english]{babel}
\usepackage[dvipsnames]{xcolor}     
\usepackage{graphicx}
\usepackage{amsmath,amssymb}
\usepackage{amsfonts}
\usepackage{mathtools}
\usepackage{braket}
\renewcommand{\eqref}[1]{(\ref{#1})}
\makeatletter
\renewcommand{\fnum@figure}{\textbf{Fig.~\thefigure}}
\makeatother

\usepackage{bm}

\usepackage{hyperref}
\hypersetup{colorlinks,citecolor={blue},linkcolor={red},urlcolor={red}}

\renewcommand{\eqref}[1]{(\ref{#1})}
\usepackage{enumitem}
\usepackage{pgfornament}
\usepackage{empheq}

\begin{document}

\title{Photon induced near-field electron microscopy from nanostructured metallic films and membranes}
\date{\today}

\author{Sophie Meuret$^*$}\email[]{sophie.meuret@cemes.fr}
\affiliation{CEMES-CNRS, Université de Toulouse, CNRS, Toulouse, France}
\author{Hugo Lourenço-Martins}
\affiliation{CEMES-CNRS, Université de Toulouse, CNRS, Toulouse, France}
\author{Sébastien Weber}
\affiliation{CEMES-CNRS, Université de Toulouse, CNRS, Toulouse, France}
\author{Florent Houdellier}
\affiliation{CEMES-CNRS, Université de Toulouse, CNRS, Toulouse, France}
\author{Arnaud Arbouet$^*$}\email[]{arnaud.arbouet@cemes.fr}
\affiliation{CEMES-CNRS, Université de Toulouse, CNRS, Toulouse, France}

\begin{abstract}
    We investigate - both experimentally and theoretically - the inelastic interaction between fast electrons and the electromagnetic field scattered by metallic apertures and nanostructures on dielectric membranes using photon induced near-field electron microscopy. The experiments - performed in a high brightness ultrafast transmission electron microscope - on gold apertures on silicon nitride membranes reveal strong modulations of the electron-light coupling strength. We demonstrates that this effect results from the combined action of the electric field scattered by the aperture edges and the reflection and transmission of the incident wave by the dielectric membrane.
    Moreover, when a nanostructure is added inside the metallic aperture, the new scattered field interferes with the previous contributions, thus imprinting the optical response of the nanostructure in additional modulations of the electron-light coupling strength. Using systematic electrodynamics simulations based on the Green dyadic method, we quantitatively analyze these different contributions to the electron-light coupling and propose further applications.
    \end{abstract}
    
\maketitle

\section{Introduction}

Ultrafast Transmission Electron Microscopes (UTEM) combining the femtosecond temporal resolution of ultrafast optical spectroscopies and the nanometric spatial resolution of electron microscopy have opened up many new possibilities to investigate light-matter at unique spatio-temporal scales \cite{zewail_4d_2009, arbouet_chapter_2018} such as the efficient probing of nano-optical excitations \cite{abajo_electron_2008, barwick_photon-induced_2009,feist_quantum_2015} or the coherent control of free electron wavefunctions \cite{echternkamp_ramsey-type_2016, priebe_attosecond_2017,pomarico_mev_2017, kfir_controlling_2020}.
Recently, the combined use of tailored illumination and inelastic electron-light interaction has been proposed to correct the spherical aberration of electron microscopes \cite{konecna_electron_2020}.
In a first proof-of-principle experiment in this direction, the electron-light coupling mediated by a dielectric membrane illuminated by a tilted plane wave has been exploited to imprint different transverse intensity profiles on the electron wave function \cite{madan_ultrafast_2022}.\\

The interaction of a tilted plane wave with a dielectric membrane or metallic mirror can also be used in so-called holographic PINEM experiments.
While conventional PINEM experiments give access to the optical-near field intensity along the electron trajectory, holographic PINEM experiments detect the interference of the studied optical excitation with a reference wave generated e.g. by a reflection of the incident wave by a planar interface.
It has been shown that such interference between the plasmon field excited at a metal/dielectric interface and the reflection from the sample can imprint the phase of the plasmon field in the electron/near-field coupling constant extracted from the PINEM signal \cite{madan_holographic_2019}. Holographic PINEM experiments have also been performed in a different geometry involving two sequential inelastic interactions induced by two distinct samples placed at different locations along the electron beam trajectory \cite{echternkamp_ramsey-type_2016}. 
Even though the presence of a membrane in electron spectroscopy experiments is unavoidable, its has been shown that it is possible to minimize its influence on the electron-light coupling by choosing the incidence angle so that the contributions from the incident and reflected electric fields almost completely cancel each other. 
This is however not always possible in particular when the space available in the objective lens of the microscope is very limited or when short focal distance focusing optics are used on the sample.
In these latter cases, a detailed knowledge of the contribution from the membrane to the inelastic signal is required prior to any demanding PINEM experiment.

\begin{center}
\begin{figure*}[htp]
\centering
\includegraphics[width=14cm,angle =0.]{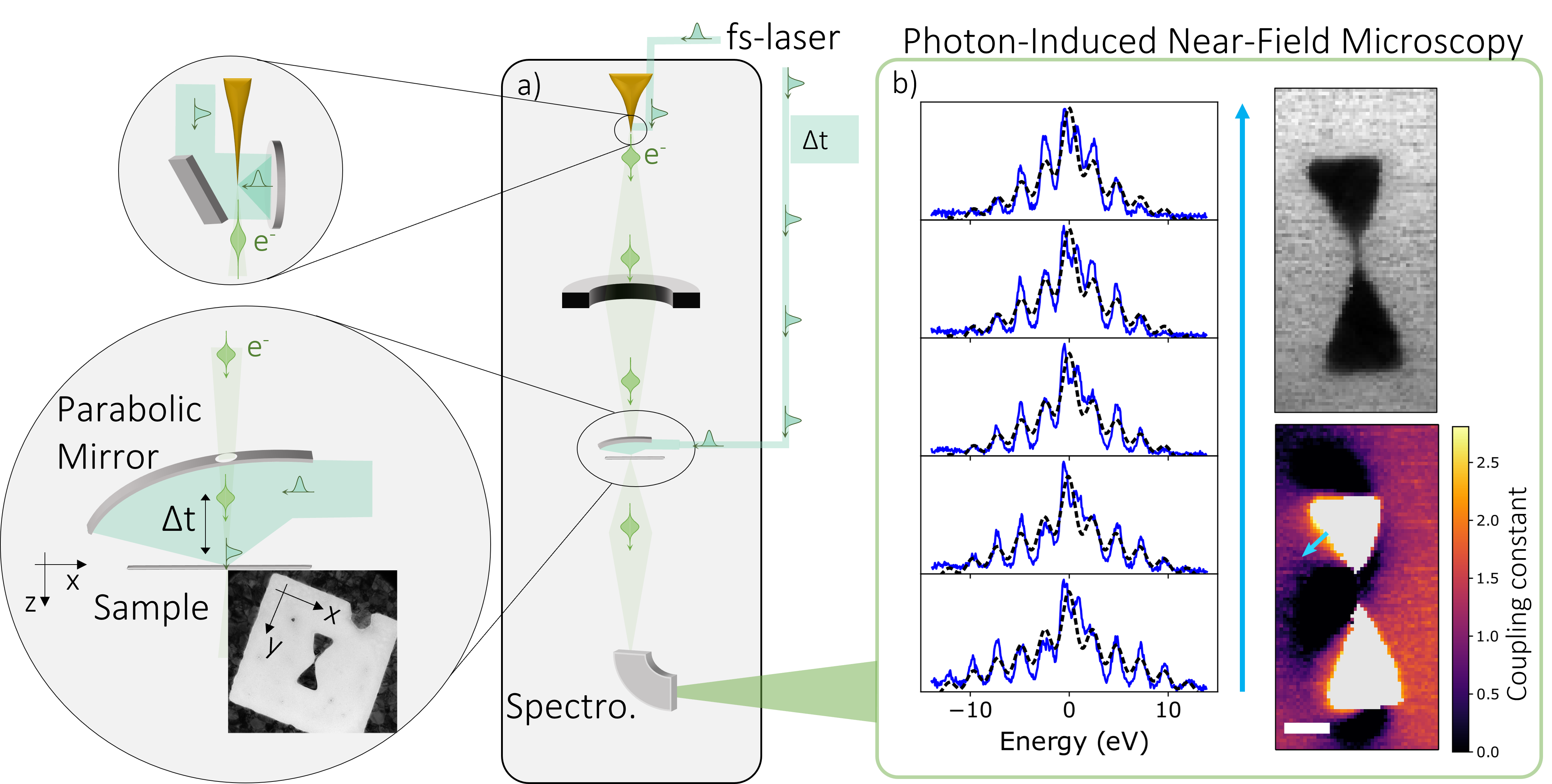}
\caption{(Color Online) a) Photon-Induced Near-field Electron Microscopy. The excitation laser pulse is focused on the sample by a parabolic mirror (incidence angle $\theta_i = 35^{\mathrm{\circ}}$). The electron and laser pulses are synchronized by a delay line (not shown) and the energy spectrum of the electron pulse is analyzed by an electron spectrometer after interaction with the optical pulse. Inset: Bright field image of the sample recorded with a continuous electron beam. b) Electron energy spectra (solid line) and fit (dashed lines) recorded at different positions along the arrow. Right Top : Bright field image recorded at the same time as the electron energy spectrum. Right bottom: Map of the electron-light coupling constant (g) extracted from a fit of the electron energy spectra acquired at each pixel.}
\label{Figure1}
\end{figure*}
\end{center}

In this study, we investigate both experimentally and theoretically the inelastic interaction between fast electrons and the electromagnetic field scattered by metallic apertures and nanostructures fabricated on a dielectric membrane (see Figure \ref{Figure1}).
Using Photon Induced Near-field Electron Microscopy, we map the inelastic interaction probability and analyze the combined influence of the scattering by the aperture edges or nanostructures and the electric field reflected or transmitted by the dielectric membrane illuminated by a tilted plane wave on the electron-light coupling.
We have performed two sets of experiments on apertures and nano-antennas fabricated in a gold film deposited on a silicon nitride membrane : first we have investigated the spatial distribution of the inelastic signal when the fast electrons travel through simple apertures of different shapes engraved in a 50 nm gold film with a focused ion beam (See SI)  before considering the case in which a gold nanostructure stands in the middle of the aperture.
The results of the PINEM experiments are analyzed using electrodynamical simulations based on the Green dyadic Method.

\section{Inelastic electron-light interactions in apertured metallic films}

PINEM works on a so-called pump-probe scheme: A first laser pulse (pump) excites the optical near-field around a nanostructure which is then probed by a subsequent electron (probe) pulse. During its transit in the optical near-field, the travelling electron can emit or absorb photons, thus leading to a modification of its energy. 
This inelastic interaction yields a characteristic electron energy spectrum composed of a series of peaks reflecting the discrete nature of the photon exchange. 
The magnitude of these peaks only depends on the so-called electron-light coupling strength $g$ which is proportional to the Fourier transform of the electric field component along the electron trajectory \cite{abajo_electron_2008, park_photon-induced_2010, feist_quantum_2015}:
\begin{equation}
    g = \frac{ e}{2 \hbar \omega} \int dz \; E_z (z) e^{- \imath \omega z/v}
    \label{definition_g_main}
    \end{equation}

    After interaction, the exit wavefunction of the electron, initially having an energy $E_0$, is a superposition of wavelets of different kinetic energies $E_n = E_0 + n \hbar\omega$.
    The amplitude of the different components is a function of the electron-light coupling constant $g$.
    
    The PINEM experiments have been performed on the high-brightness ultrafast TEM developed in CEMES-CNRS.
    The latter is a customized 200 kV cold-field emission Hitachi High-Technology HF2000 \cite{houdellier_development_2018}.
    The electron gun has been modified so that femtosecond laser pulses can be focused onto the tungsten nanotip and trigger the emission of femtosecond electron pulses \cite{caruso_development_2017}.
    The high brightness of the femtosecond electron source provides sub-400 fs electron pulses that can be focused in spots as small as 1 nm on the sample.
    The electron microscope has been modified to allow optical excitation of the sample inside the objective lens.
    A high numerical aperture parabolic mirror with XYZ translation stage has been added between the objective lens pole pieces yielding a tilt angle of $35^{\circ}$ between the electron probe and the optical pump focused on the sample (see Figure \ref{Figure1}).
    Figure \ref{Figure1}-b shows an example of the nanostructures investigated by PINEM in this study.
    These structures are apertures and nano-antennas fabricated from a 40 nm thick gold film evaporated on a 50 nm thick $\mathrm{Si_{3}N_{4}}$ membrane (see Methods for more details). 
    At each point of the map, an electron energy spectrum such as the ones shown in Figure \ref{Figure1}-b is acquired with a Gatan PEELS 666 with a  typical integration time between $300$ and $500$ ms/pixel.
    The collection of energy spectra is then post-processed to extract the electron-light coupling constant at each point of the map (see Methods for more details about PINEM theory and data processing).
    In the experiments reported in this study, the electron beam is accelerated at 150 keV.
    The use of a 30 $\mu$m STEM aperture allows to focus the electron probe in a few nm spot \cite{houdellier_development_2018, caruso_high_2019}.
    The geometry of the experiments is sketched in Figure 1. The electron is travelling along the (Oz) axis, perpendicular to the (OXY) plane of the membrane.
    The position of the electron beam on the membrane is identified by the coordinates x and y. 
    The (Ox) axis lies in the plane of incidence whereas the (Oy) axis is perpendicular to the latter.    

\begin{center}
    \begin{figure}[htp]
    \centering
    \includegraphics[width=\columnwidth]{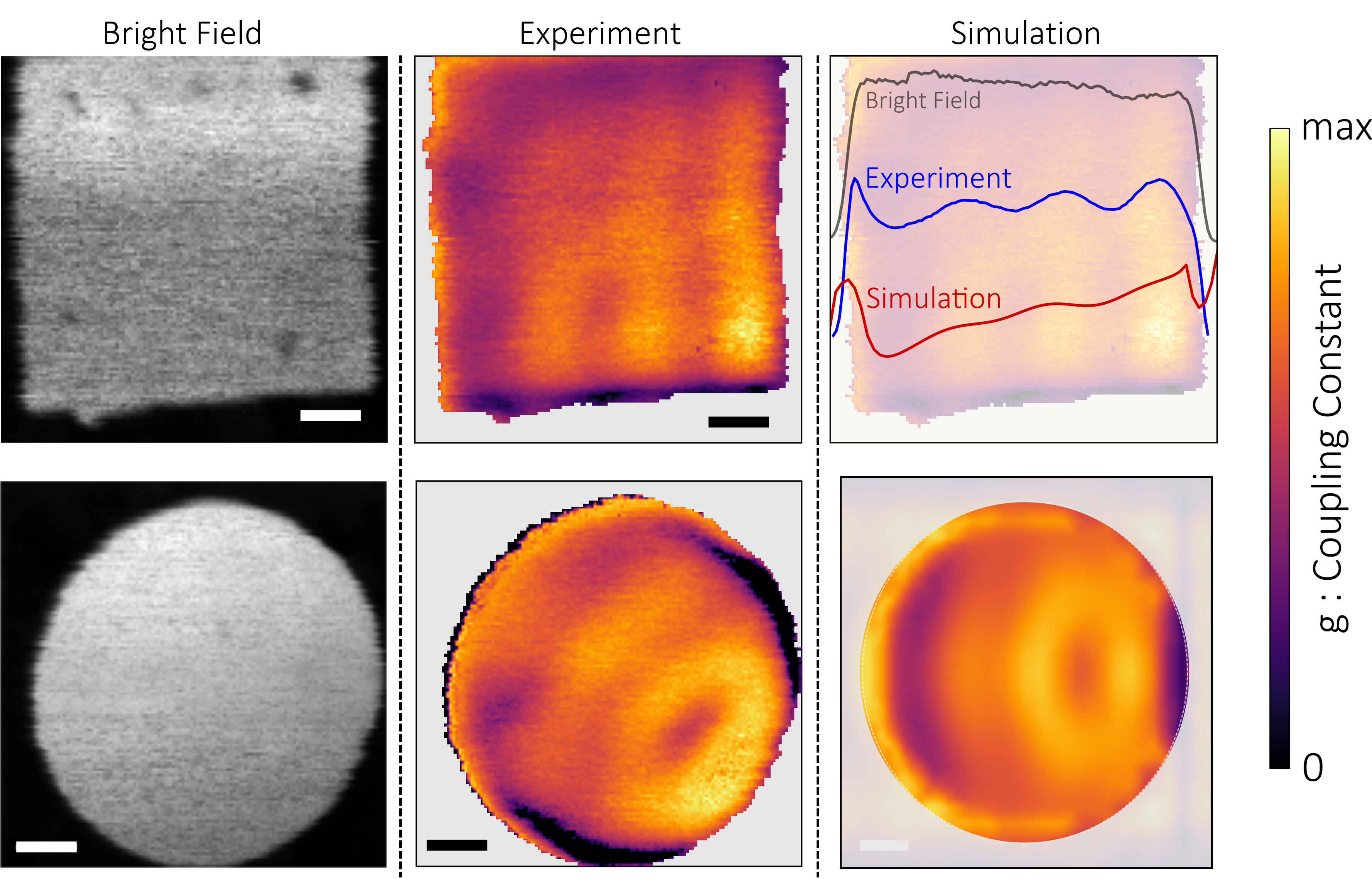}
    \caption{(Color Online) Map of the electron-light coupling strength on an apertured gold film deposited on a silicon nitride membrane.
    Bright field image of respectively a square (top) and a circle (bottom) aperture fabricated in a 40 nm thick gold film deposited on a 50 nm thick $\mathrm{Si_{3}N_{4}}$ membrane.
    Experiment : Map of the electron-light coupling strength $g$ extracted from numerical fits of the electron energy spectra for the square (top) and circle (bottom) aperture.
    Simulation: Electron-light coupling constant computed from electrodynamical simulations based on the Green Dyadic Method. Top : Overlay on top of the experimental data, a comparison of the simulation (cyan line) and experiment (dark blue line), the gray dashed line represent the bright field intensity profil). Bottom : 2D simulation map in the case of the circle aperture}
    \label{Figure2}
    \end{figure}
    \end{center}

In a first set of experiments, we study the electron energy exchanges when the fast electrons travel through apertures of different shapes.
The spatial variations of the electron-light coupling constant $g$ extracted from the PINEM maps acquired on square and circular apertures are shown in Figure \ref{Figure2}.
Clear modulations of $g$ are visible in the apertures.

The origin of these modulations can be traced back from the definition of the coupling constant $g$ (Equ. (\ref{definition_g_main})) that governs the electron-light interaction probability.
In vacuum, the momentum mismatch between free space light and the moving electron leads to a vanishing $g$ and therefore no interaction.
We provide in supplementary information additional data acquired on similar apertures without the silicon nitride membrane confirming the absence of detectable inelastic signal at the center of the aperture.
The presence of a membrane on the electron path leads to a non null integral and therefore to inelastic interaction probabilities that depend on the wavelength, intensity and angle of incidence of the incident wave as well as the membrane dielectric constant \cite{vanacore_attosecond_2018}.
Taking into account the multiple reflections at the two vacuum/membrane interfaces,  $g_{mem}$ can be written as:
\begin{equation}
g_{mem} =  \frac{ \imath \, e}{2 \hbar \omega} e^{\imath k_{i,x}.x}  f ( \omega, \theta_i, n_m, v)
\label{def_gmem_main}
\end{equation}
in which $f$ is a complex function of the angular frequency of light $\omega$, the electron speed $v$, the membrane refractive index $n_m$ and thickness $d$ and the incidence angle $\theta_i$.
The complete expression of $g_{mem}$ taking into account the multiple reflections at the two vacuum/membrane interfaces is given in supplementary information.
It is clear from equation (\ref{def_gmem_main}) that the modulus of $g_{mem}$ on a simple membrane is not expected to display any spatial modulation.

\begin{center}
    \begin{figure*}[htp]
    \centering
    \includegraphics[width=16cm,angle =0.]{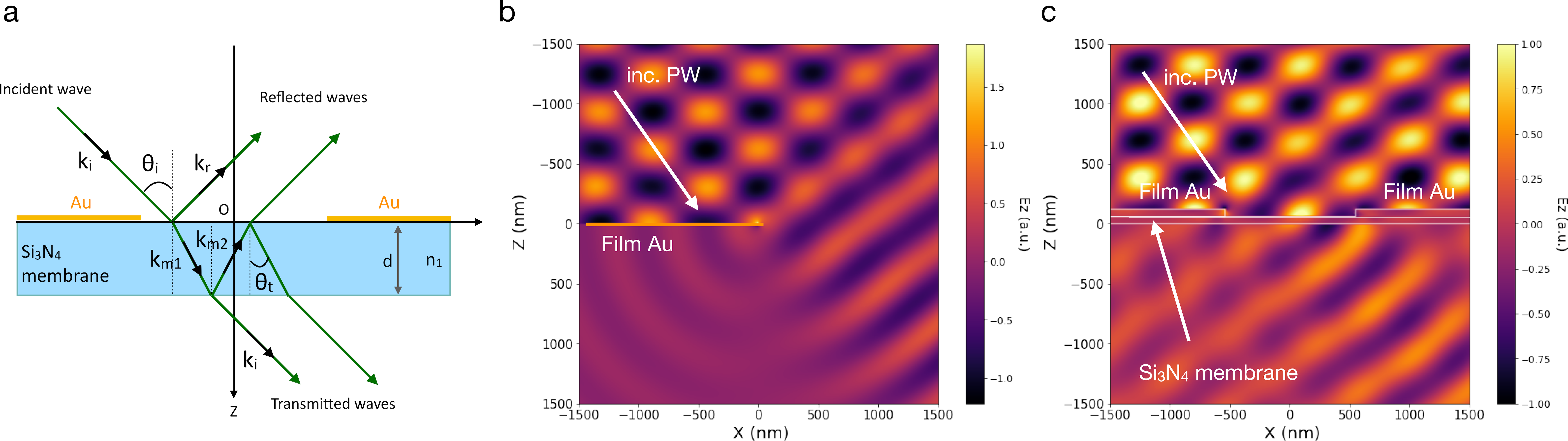}
    \caption{(Color Online) a) Sketch of the experiment showing the beams reflected and transmitted by the membrane. b) Component along the electron trajectory of the electric field scattered by a semi-infinite half plane as predicted by the Sommerfeld model. c) Total electric field on the sample computed from 2D electrodynamical simulations based on the Green Dyadic Method.}
    \label{Figure3}
    \end{figure*}
    \end{center}

The spatial modulations of the electron-light coupling strength visible in Figure (\ref{Figure2}) may arise from the combined influence of (i) the contrast in reflectivities between the metal and the dielectric surface and 
(ii) the scattering from the film edge.
First, the existence of a separation between two regions of different reflectivities in the sample plane makes the moving charge interact with the electric field reflected either by the gold surface or the silicon nitride membrane depending on the distance between the electron and the membrane.
The transition between the two cases does not occur for the same value of $z$ depending on the distance of the electron beam to the aperture edge.
The expression of the electron-light coupling strength resulting solely from the difference in reflectivities predicted by a simple model in the geometrical approximation can be found in the SI.
This model predicts a spatial modulation of the electron-light coupling strength in poor agreement with the experiment.
The origin of the discrepancy lies in the fact that a crude geometrical approach neglects the important contribution of the wave scattered by the film edge.
For p-polarized illumination, the scattering by the film edge yields an electric field having a z-component comparable to the incident wave and therefore a significant contribution to the electron-light coupling.
The latter is expected for p-polarized illumination to contribute to the electric field along the electron trajectory.
The electric field scattered by the metallic film edge predicted by the Sommerfeld model is shown in Figure \ref{Figure3}-b \cite{BornWolf:1999:Book}.

To take into account the different contributions discussed above, we have performed electrodynamical simulations using the Green Dyadic Method (GDM) \cite{martin_generalized_1995, girard_near_2005}.
We have used the pyGDM open-source python toolkit \cite{wiecha_pygdm_2022}. pyGDM relies on the concept of a generalized propagator and allows to perform electro-dynamical simulations giving access to a large variety of near-field or far-field optical properties of individual nano-structures under either optical or electronic excitation \cite{arbouet_electron_2014}.
More details about the GDM simulations are provided in the supplementary information.
The influence of the scattering by the metallic film edge and the difference in reflectivities from the gold film and silicon nitride membrane have first been simulated using 2D GDM simulations considering a silicon membrane half covered by a gold film.
We have plotted in Figure \ref{Figure2}-c the electron-light coupling strength computed using 2D-GDM across a 1100 nm gap made in a gold film deposited on a silicon nitride membrane.
The results of the 2D-GDM calculations are in good agreement with the profile extracted from the experimental data acquired on the square aperture.
We show in Figure \ref{Figure3}-c the total electric field on the sample computed using 2D-GDM.
A detailed study of the different contributions to the spatial variations of the electron-light coupling constant in the aperture provided in the supplementary information shows that scattering by the aperture edges and the reflection/transmission by the dielectric membrane contribute in comparable proportions to the modulation of the inelastic interaction strength.
The case of more complex in-plane shapes requires full 3D electrodynamical simulations.
Figure \ref{Figure2}-f shows the results of 3D-GDM calculations performed on the circular aperture.
Again, a good agreement is obtained between the experiment and the electrodynamical calculations.

\begin{center}
    \begin{figure*}[htp]
    \centering
    \includegraphics[width=14cm,angle =0.]{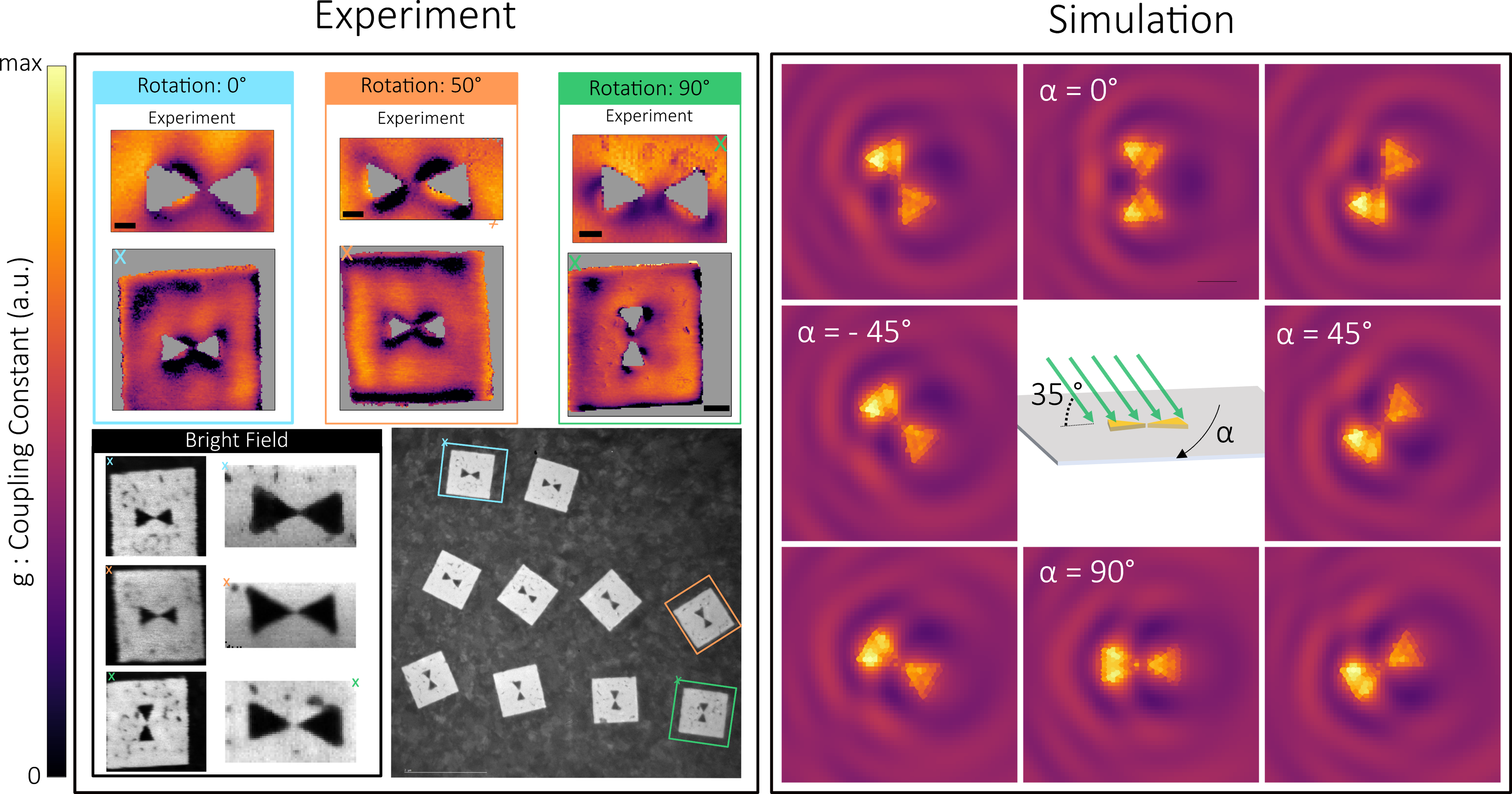}
    \caption{(Color Online) a) Bright field image of the studied sample.
    b-d) PINEM maps of the electron-light coupling constant $|g|$ acquired on 3 different apertures having different orientations with respect to the illumination.
    Maps of the electron-light coupling constant $|g|$ computed using 3D-GDM simulations on a gold bow-tie (200 nm edge length, 160 nm gap) placed in vacuum (e) or on a silicon nitride membrane (f).}
    \label{Figure4}
    \end{figure*}
    \end{center}

In a second set of experiments, we have studied the electron-light coupling strength on apertured metallic films in which a nanostructure has been added on the silicon nitride membrane in the square apertures.
As shown in Figure \ref{Figure4}-a, the sample consists in square apertures with a bowtie antenna made of two equilateral gold prisms with an edge length $e = 200$ nm located at the center of the aperture.
The exact same geometry has been fabricated several times with varying orientations with respect to the optical excitation.
Figure \ref{Figure4} shows the bright field images and maps of the electron-light coupling strength extracted from the electron energy spectra acquired at each position.
Away from the bowtie antenna, the spatial distribution of the inelastic interaction strength is very similar to the case of the empty aperture.
Closer to the nano-antenna, the interaction strength shows complex spatial variations with both regions in which the presence of the nano-antenna reinforces the coupling strength and regions in which the latter is diminished with respect to the case of the empty aperture.
To analyze these observations, we have performed 3D-GDM simulations considering either a gold bowtie in vacuum or a gold bowtie on a silicon nitride membrane.
The results are displayed in Figure \ref{Figure3}-e and f. 
When the gold bow-tie is in vacuum, the electron-light coupling vanishes away from the antenna : the momentum mismatch between the fast particle and free space light prevents their coupling outside of the near-field zone.
In the near-field region, the spectrum of the optical near-field includes large wavevector components that couple efficiently with the moving electron.
When the orientation of the particle with respect to the plane of incidence is changed, the regions where the electron couples efficiently to the electron are also modified, following the topography of the optical near-field of the nano-antenna.
In the case of a nano-antenna on a membrane, the situation is more complex.
The electron-light coupling does not vanish away from the antenna as the dielectric contrast at the surface of the substrate mediates the coupling between the incident wave and the moving charge.
The measured electron-light coupling then results from the combined influence of the membrane and the gold nano-antenna.
Neglecting the mutual influence of the membrane and nano-antenna, the electron-light coupling can be written as : 

\begin{equation}
    g_{tot} =   \frac{ e}{2 \hbar \omega} \int dz \left [ E_z^{mem} (z) +  E_z^{ant} (z) \right] \, e^{- \imath \omega z/v}
    \label{gtot}
    \end{equation}
    
    Equation (\ref{gtot}) shows that the regions in which the electron-light coupling either increased or decreased is clearly visible on the experimental results of Figure \ref{Figure4}-b-d originate from the interference between the optical response of the membrane and the antenna.
    In addition to the electric field of the nanostructure itself, the use of a p-polarized tilted illumination in our PINEM experiments yields electric fields reflected/transmitted by the membrane or scattered by the aperture edges that have a non null component along the electron trajectory and efficiently couple with the moving charge.
    These electric fields can interfere with the optical field scattered by the illuminated nano-objects.
    The experimental and theoretical results of Figure \ref{Figure4} reveal these interferences in the near-field of gold nano-antennas but similar interference effect should be visible in the far-field of a nano-scatterer.

\begin{center}
    \begin{figure*}[htp]
    \centering
    \includegraphics[width=16cm,angle =0.]{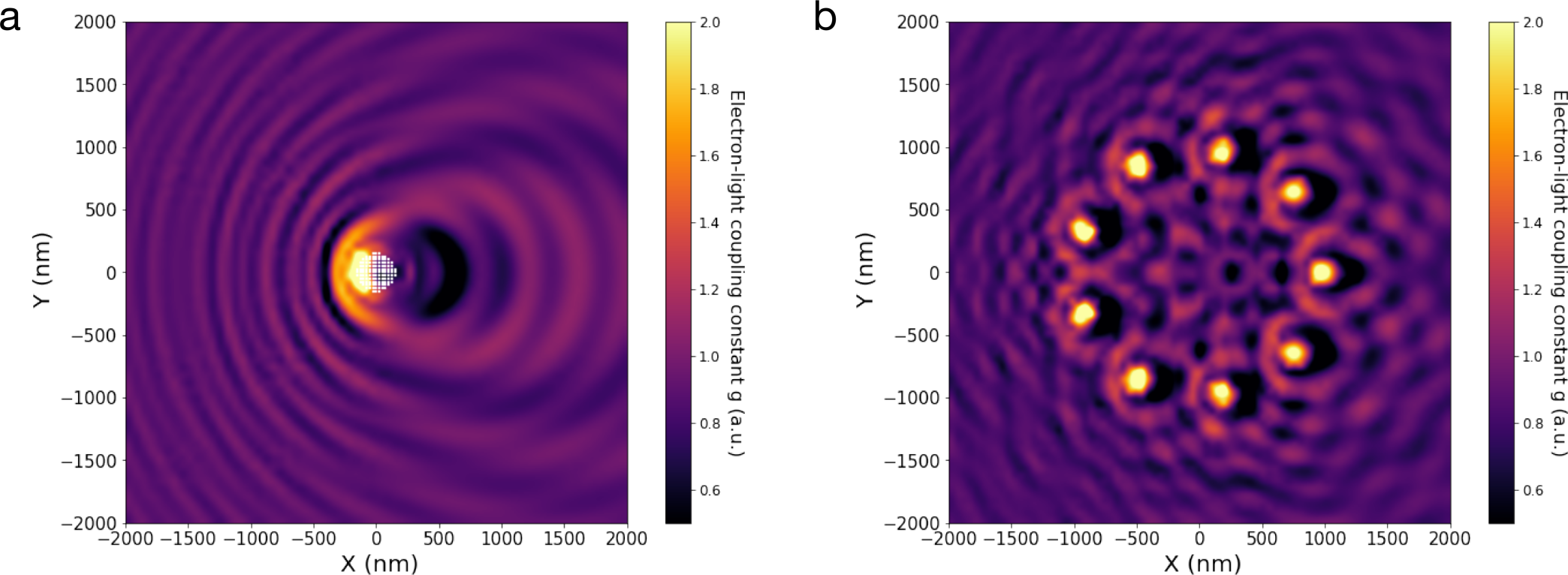}
    \caption{(Color Online) a) PINEM map computed on a $2 \mu\mathrm{m} \times 2 \mu\mathrm{m}$ centered on a gold disc ($R = 150$ nm, $H = 300$ nm) on a silicon nitride substrate.
    b) PINEM map computed on an optical corral composed of 10 identical gold nanodiscs  ($R = 50$ nm, $H = 300$ nm) on a silicon nitride substrate.}
    \label{Figure5}
    \end{figure*}
    \end{center}
    To address this point, we simulated PINEM experiments on individual nano-scatterers deposited on a membrane and illuminated by a tilted plane-wave ($\theta_i = 35^{\circ}$).
    The nano-scatterers were chosen high enough so that the tilted illumination yields an out-of-plane dipole capable of radiating electric fields with z-components of magnitude large enough to interfere efficiently with the incident wave.
    Figure \ref{Figure5}-a displays the results in the case of a single gold nanodisc.
    The electron-light coupling strength displays spatial modulations reminiscent of the Doppler effect arising from the interference between the tilted incident plane wave and the secondary wave scattered by the excited metallic nano-object.
    The interferences obtained here between the scattering from the nanodiscs and the reflection from the dielectric substrate are connected to the results of \cite{madan_holographic_2019} where the interferences between the surface plasmon polariton field propagating away from an optically excited nano-hole and the reflection from the sample were reported.
    Contrary to the case of Figure \ref{Figure4} these interferences appear in the far-field zone, the presence of the dielectric contrast at the membrane surface allowing the coupling between the electron and the electromagnetic field.
    As illustrated in Figure \ref{Figure5}-b, when more nanostructures are considered, the spatial modulation of the electron-light coupling results from the interference between (i) the incident wave, (ii) the waves reflected/transmitted by the substrate and (iii) the  different waves scattered by the nanostructures.
    In the case of PINEM performed using a normally incident illumination, the spatial distribution of the electron-light coupling resembles closely the topography of the amplitude of the z-component
    of the total electric field.
    The case of a tilted illumination is associated with a significant contribution from the substrate to the electron-light coupling that yields more complex modulations of the inelastic interaction probability.
    Our results show that experiments performed in this configuration demand careful electrodynamical simulation to take into account the different contributions to the electromagnetic field probed by the moving electron.
    
    \section{Conclusion}

    In conclusion we have studied both experimentally and theoretically the electron-light coupling in apertures and nanostructures fabricated on a dielectric membrane.
    Our results show that the scattering from the aperture edges as well as the electric field reflected or transmitted by the dielectric membrane illuminated by a tilted plane wave contribute significantly to the modulation of the electron-light coupling strength measured in PINEM.
    This contribution from the membrane  interferes with the electric field scattered by nanostructures fabricated on the membrane and alters the electron-light coupling both in the near-field and far-field region.
    Whereas the analysis of the experimental results in this configuration requires a careful analysis and comparison with electrodynamical simulations, these interference effects could be exploited to map the phase of the electric fields scattered by complex nanostructures.

    \section{Acknowledgements}
    
    This project has been funded in part by the European Union’s Horizon 2020 research and innovation program under Grant Agreement Nos. 823717 (ESTEEM3).
    This project has been funded in part by the ANR under Grant Agreement No. ANR-19-CE30-0008 ECHOMELO and Grant Agreement No. ANR-14-CE26-0013 FemtoTEM.
    This work was supported by Programme Investissements d'Avenir under the Program ANR-11-IDEX-0002-02, reference ANR-10-LABX-0037-NEXT (MUSE grant).
    All authors declare no competing interest.

\newpage
\bibliographystyle{unsrt}

\newpage
\onecolumngrid

\section{Supplementary Information}

\subsection{Sample fabrication}
All the sample presented in this article were prepared in the same fashion. A 50 nm gold thick layer (80 nm for the data of \ref{Figure1}) was thermally evaporated at the surface of a 50 nm thick $Si_{3}N_{4}$ membrane. The gold was removed at very specific positions with a Gallium ion beam. Features as small as 10 nm were observed. 

\subsection{Electron energy gain experiments and data processing}

The fitting procedure is based on the theory developed in \cite{ feist_quantum_2015} where the probability for the electron to lose or gain $\hbar\omega$ is 

\begin{equation}
P_{n} = J_{n}(2|g|^{2})
\label{bessel}
\end{equation}

\noindent with $g$ the coupling constant described in the main text and $J_{n}$ the Bessel function of first order. Because of the spatial and temporal spread of our electron probe and the duration of the laser pulse (150 fs) compared to our electron pulse (300fs) the electron-light coupling strength $g$ measured experimentally is the average of different values that vary in space and time. Therefore, we consider a gaussian distribution of the $g$ values centered around $g$ and with a spread of $\delta g$. The final fitting function is therefore

\begin{subequations}
    \begin{empheq}[left={\empheqlbrace\,}]{align}
    \Gamma(E)&=\sum_{n}J_{n}(2|f(g,\delta g)|)^{2}\delta(|E|-n\hbar\omega)\\
    f(g,\delta g)&=\frac{1}{2\pi\delta g}e^{-\frac{g^2}{2\delta g^{2}}}
    \end{empheq}
\label{bessel}
\end{subequations}

\noindent Moreover, to increase the signal we used the UTEM in a regime where much more than 1 electron was produced at the tip. Due to Coulomb repulsion, the chirp of the electron pulse i.e the variation of the electron energy within the duration of the electron pulse is visible in the experimental data. 
This  well-known effect leads to a splitting of the zero-loss peak.

\begin{center}
\begin{figure*}[htp]
\centering
\includegraphics[width=14cm,angle =0.]{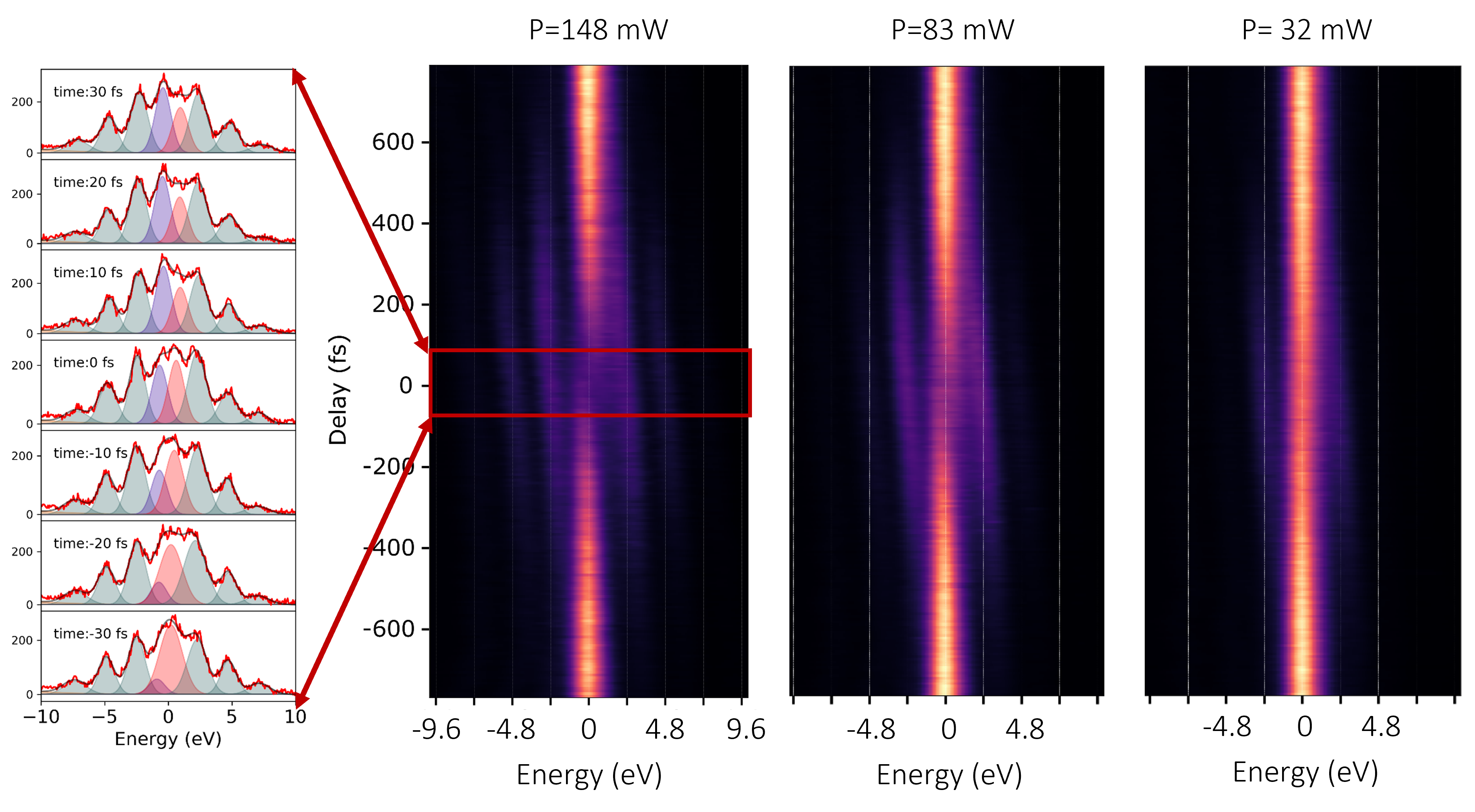}
\caption{(Color Online) EEGS spectrum vs delay between the pump and the probe at a fixed position on the sample (here close to the edge of a triangle). The chirp induces a splitting of the beam at zero delay and increases with the power of the pump laser (P=148 mW to P=32 mW here).}
\label{FigureS5}
\end{figure*}
\end{center}

\begin{center}
    \begin{figure}[h!]
    \centering
    \includegraphics[width=14cm,angle =0.]{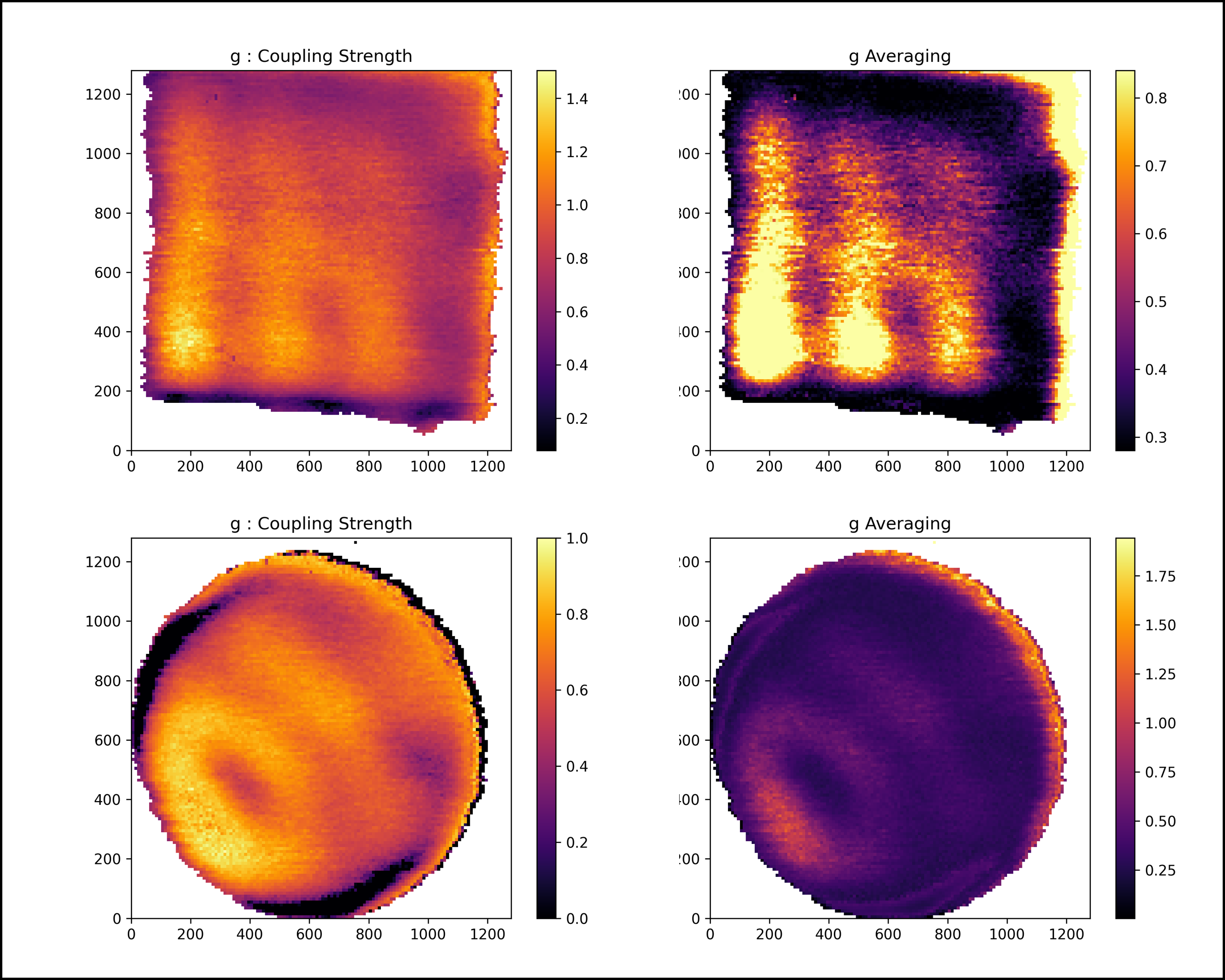}
    \caption{(Color Online)Values of the $g$ and $dg$ parameter extracted from the fitting procedure of the data presented in Figure \ref{Figure2} of the main text.}
    \label{FigureS6}
    \end{figure}
    \end{center}

\begin{center}
    \begin{figure}[h!]
    \centering
    \includegraphics[width=10cm,angle =0.]{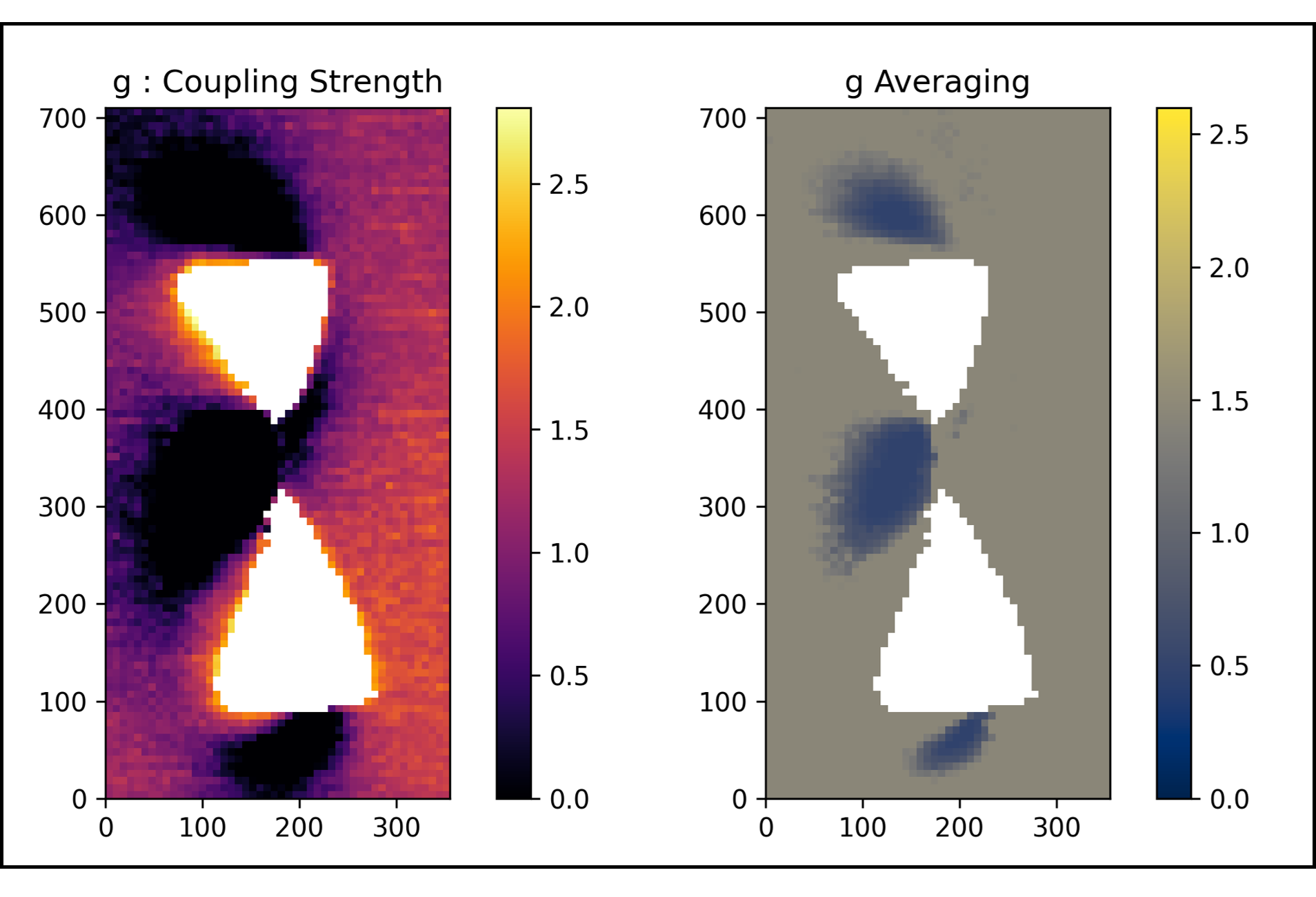}
    \caption{(Color Online) Values of the $g$ and $dg$ parameter extracted from the fitting procedure of the data presented in Figure \ref{Figure1} of the main text.}
    \label{FigureS7}
    \end{figure}
    \end{center}

\begin{center}
    \begin{figure}[h!]
    \centering
    \includegraphics[width=14cm,angle =0.]{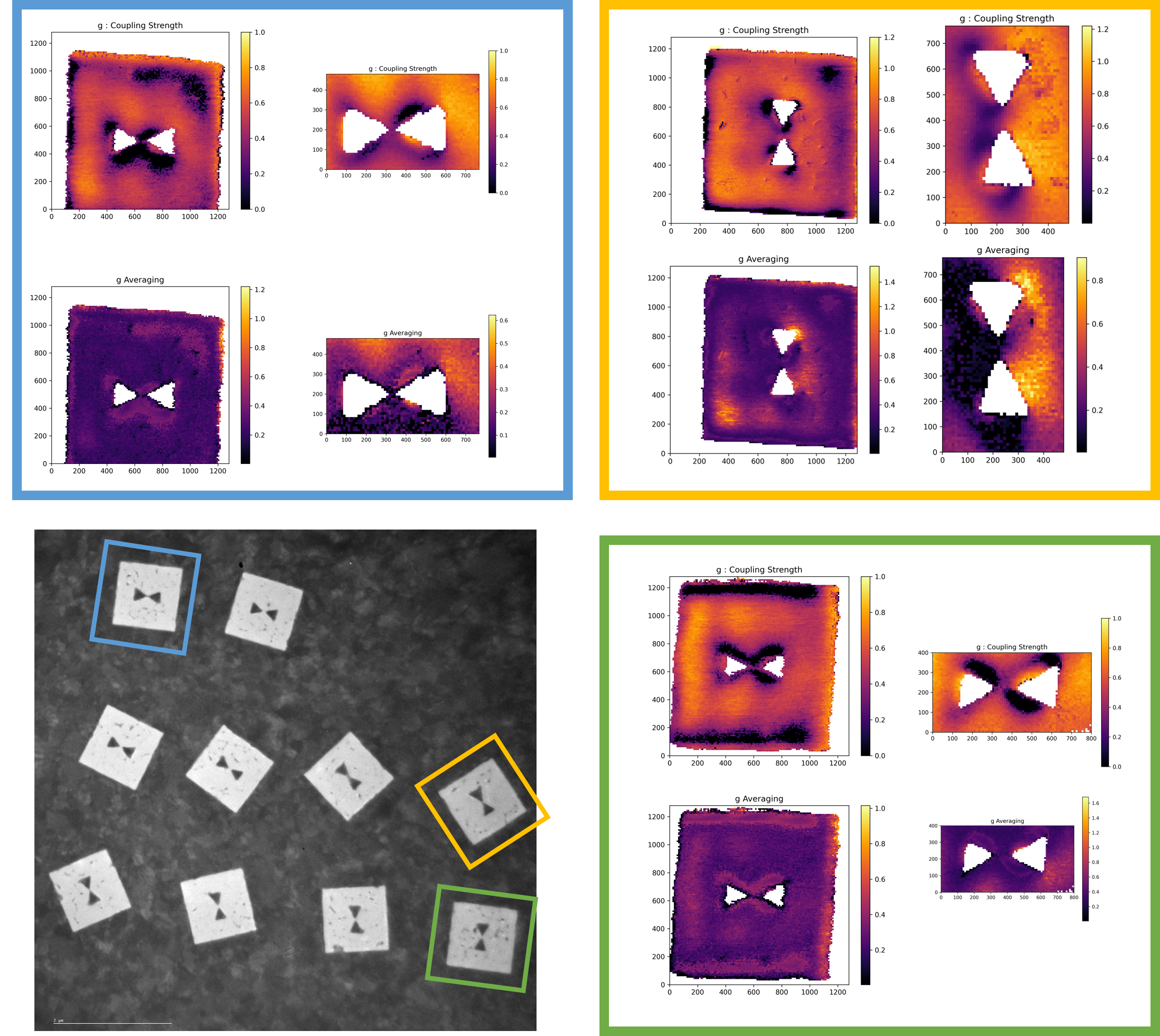}
    \caption{(Color Online)Values of the $g$ and $dg$ parameter extracted from the fitting procedure of the data presented in Figure \ref{Figure4} of the main text.}
    \label{FigureS8}
    \end{figure}
    \end{center}

\noindent This splitting depends strongly on the  value of $g$, and is therefore difficult to take into account in our fitting procedure. The double peak leads to an overestimation of $\delta g$ by the numerical fit. However, it does not change the spatial variations of $g$  but the values of g and $\delta g$ are not quantitative.\\
\noindent Fitting maps of the data of the main text with the $g$ and $\delta g$ value are shown below on Figures \ref{FigureS6},\ref{FigureS7},\ref{FigureS8}

    \subsection{PINEM experiments : complementary data}

\begin{center}
    \begin{figure}[htp]
    \centering
    \includegraphics[width=11cm,angle =0.]{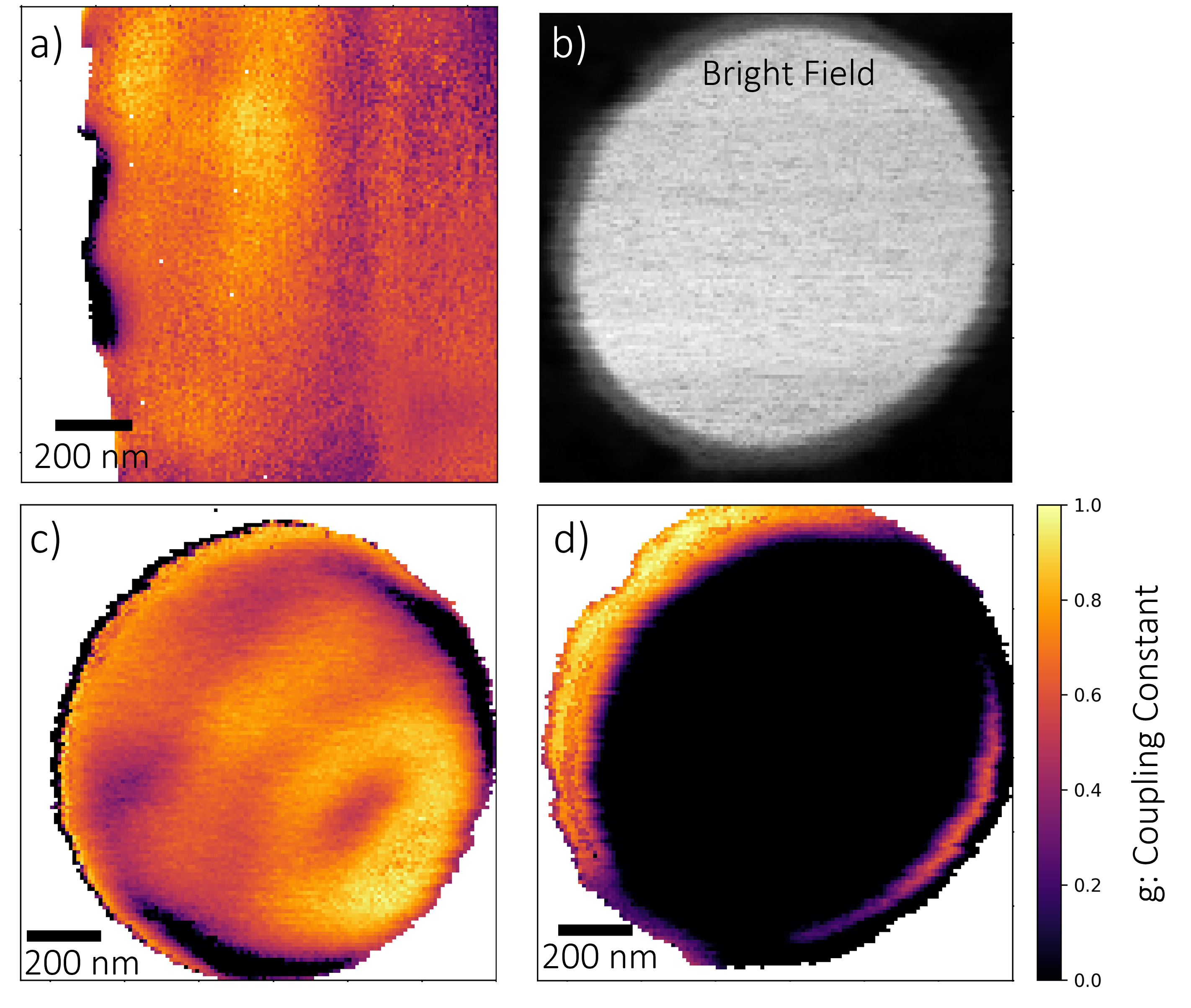}
    \caption{(Color Online) a) PINEM map in the vicinity of the edge of a 40 nm gold film  deposited on a silicon nitride membrane.
    b) Bright field image of a circular aperture similar as the one shown in Figure 2 of the main text but in which the silicon nitride membrane is absent.
    c) PINEM map from Figure 2 of the main text.
    d) PINEM map of the aperture shown in b. }
    \label{FigureS1}
    \end{figure}
    \end{center}

    \noindent We provide in \ref{FigureS1} additional experimental data on a film edge and an apertured metallic film in which the membrane has been removed from the aperture.
    No PINEM signal is detected in the center of the aperture sufficiently far from the metallic film edge whenever the membrane has been removed.
    
    \subsection{PINEM theory in a nutshell}

    The energy exchanges experienced by a fast electron can be quantified through the electron-light coupling parameter $g$.
    This parameter is proportional to the Fourier transform of the electric field component  along the electron trajectory \cite{abajo_electron_2008, park_photon-induced_2010}:
    \begin{equation}
    g = \frac{ e}{2 \hbar \omega} \int dz E_z (z) e^{- \imath \omega z/v}
    \label{definition_g}
    \end{equation}
    
    \noindent After interaction, the exit wavefunction of the electron, initially having an energy $E_0$, is a superposition of wavelets of different kinetic energies $E_n = E_0 + n \hbar\omega$.
    The amplitude of the different components is a function of the electron-light coupling constant $g$.

    \subsection{Energy exchanges induced by a perfect mirror}

    A first approximation of $g$ can be obtained in the case of a perfect mirror \cite{vanacore_attosecond_2018}.
    We assume that the electron is travelling along the (Oz) axis from negative to positive $z$.
    The sample is illuminated from the top ($z<0$).
    In this case, the z-component of the electric field above the interface can be written $E_z (z) = \left( E_{i,z} e^{\imath {\bf k_i }.{\bf r }} + E_{r,z} e^{\imath {\bf k_r }.{\bf r }}  \right) f(z) H(-z) $ in which $H$ is the Heaviside step function
    and $f(z)$ is the envelope of the laser pulse.
    We assume that $f(z)$ is a gaussian function of temporal width $\sigma_t$ :
    $f(z) = e^{-z^2/2 v^2 \sigma_t^2}$.
    Assuming that $1/v \sigma_t < < |\omega/v - k_i |$, Equ.(\ref{definition_g}) yields the following expression:
    $$
    g = \frac{\imath \, e}{2 \hbar \omega} \left\{  \frac{E_{i,z}}{\omega/v - k_{i,z} } + \frac{E_{r,z}}{\omega/v - k_{r,z} }   \right\} e^{\imath k_{i,x}.x} 
    $$
    with $k_{i,z} =  k \cos \theta_i$, $k_{r,z} = - k \cos \theta_i $ and $k_{i,x} = k \sin \theta_i = k_{r,x}$, $\theta_i$ being the incidence angle.
    
    \subsection{Energy exchanges induced by a dielectric membrane}    

\begin{center}
    \begin{figure}[htp]
    \centering
    \includegraphics[width=14cm,angle =0.]{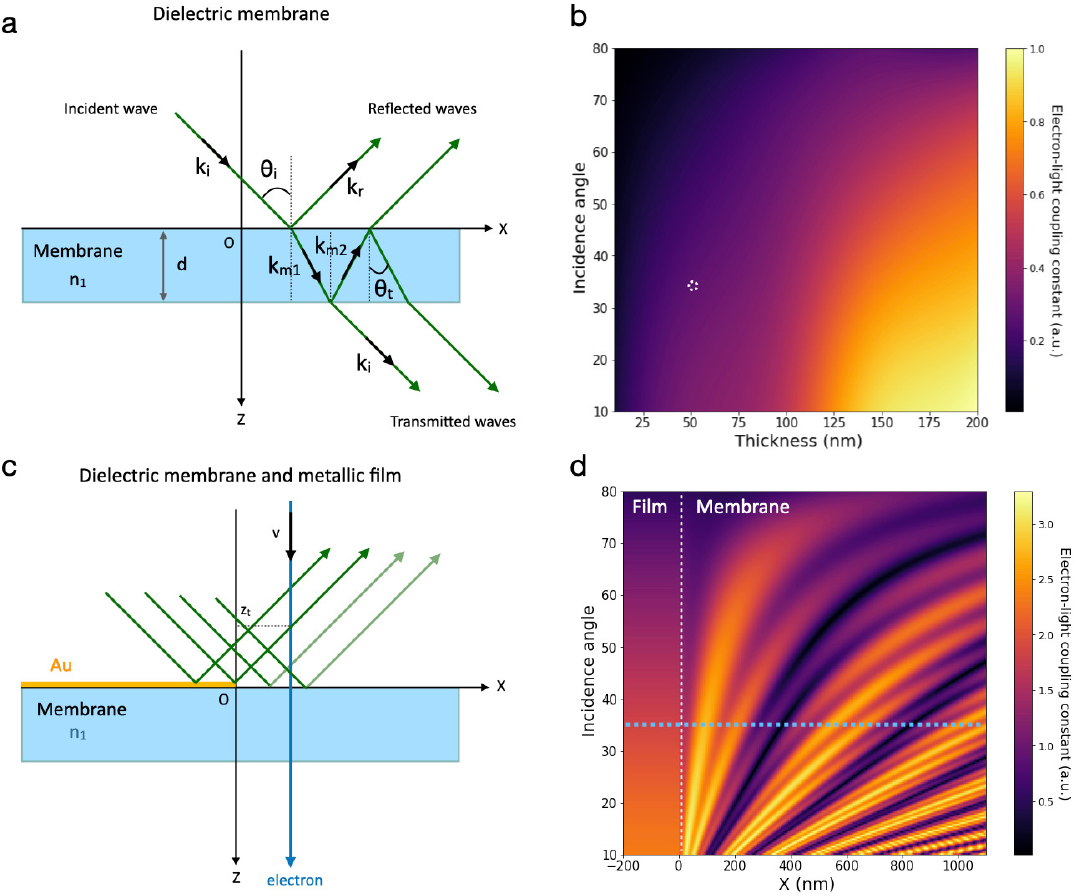}
    \caption{(Color Online) a) Reflection and transmission of a plane wave by a dielectric membrane. b) Electron-light coupling constant computed at $\lambda_{inc} = 515$ nm as a function of the membrane thickness and incidence angle. 
    c) Reflection and transmission of a plane wave by a dielectric membrane half covered bya metallic film.
    d) Electron-light coupling constant computed at $\lambda_{inc} = 515$ nm as a function of the incidence angle and distance to the edge of the metallic film. }
    \label{FigureS2}
    \end{figure}
    \end{center}

    In the case of a membrane of thickness $d$ and refractive index $n_m$, a more complex expression of $g$ taking into account all the reflections at the membrane interface can be obtained.
    The considered situation is depicted in Figure S1-a.
    The membrane interfaces are located at $z=0$ and $z = d$.
    We still consider an illumination from the top ($z < 0$) and only consider p-polarized illumination as only the z-component of the electric field acts on the moving electron.
    The amplitude of the different reflected and transmitted waves will be given by the Fresnel amplitude reflection (resp. transmission) coefficients $r_{12}$ (resp. $t_{12}$).
    For a wave incident from a medium of real refractive index $n_1$ to a medium of real refractive index $n_2$ these coefficients can be written as :
    \begin{eqnarray}
    r_{12, p} = \frac{E_{1r,p}}{E_{1i,p}} = \frac{n_2 \cos \theta_i - n_1 \cos \theta_t }{n_2 \cos \theta_i + n_1 \cos \theta_t} \\
    t_{12, p} = \frac{E_{2,p}}{E_{1i,p}} = \frac{2 n_1 \cos \theta_i  }{n_2 \cos \theta_i + n_1 \cos \theta_t}
    \end{eqnarray}
    $\theta_i$ and $\theta_t$ are the angle of incidence (resp. refraction) at the 1/2 interface.
    
    \noindent Above the membrane, the electric field is the sum of the incident field and all reflected waves from the interface:
    \begin{eqnarray*}
    E_{I, p} ({\bf r}) &=& e^{\imath \, {\bf k_i} . {\bf r} } +   e^{\imath \, {\bf k_r} . {\bf r} } \left[
     r_{12} + \sum_{n=1}^{\infty} t_{12}t_{21}(r_{21})^{2n-1} e^{\imath \Phi_m n} e^{- \imath n \, {\bf k_r} . \Delta {\bf r}}\right] \\
     &=& e^{\imath \, {\bf k_i} . {\bf r} } +   a_1 ( \theta_i, n_m, d) e^{\imath \, {\bf k_r} . {\bf r} } 
    \end{eqnarray*}
    
    \noindent in which we define $a_1 ( \theta_i, n_m, d)$ as:
    \begin{equation}
    a_1 ( \theta_i, n_m, d) =  r_{12} + \sum_{n=1}^{\infty} t_{12}t_{21}(r_{21})^{2n-1} e^{\imath \Phi_m n} e^{- \imath n \, {\bf k_r} . \Delta {\bf r}}
    \end{equation}
    
    \noindent $\Phi_m$ is the dephasing induced by one round trip of the wave in the dielectric membrane.
    It can be written as:
    
    \begin{equation}
    \Phi_m = \frac{4 \pi d n_m}{\lambda \cos \theta_r } 
    \end{equation}
    $\Delta {\bf r} = (2 d \tan \theta_r, 0, 0) $ is the shift of the exit point of the wave upon one round trip inside the membrane. Inside the membrane, the electric field is the superposition of the waves reflected at the top and bottom interfaces:
    
    \begin{eqnarray*}
    E_{II, p} ({\bf r}) &=& e^{\imath \, {\bf k_{m,1}} . {\bf r}}  \left[
    t_{12} \sum_{n=0}^{\infty}(r_{21})^{2n} e^{\imath \Phi_m n} e^{- \imath n \, {\bf k_{m,1}} .  \Delta {\bf r}} \right] \\
    && +
    e^{\imath \, {\bf k_{m,2}} . {\bf r} }  \left[
     t_{12} \sum_{n=0}^{\infty}(r_{21})^{2n+1} e^{\imath \Phi_m(n+\frac{1}{2})} e^{ - \imath \, {\bf k_{m,2}} .( \Delta {\bf r}_m  + (n+\frac{1}{2}) \Delta {\bf r})}  \right]  \\
     &=& a_{m,1} ( \theta_i, n_m, d) e^{\imath \, {\bf k_{m,1}} . {\bf r}} + a_{m,2} ( \theta_i, n_m, d) e^{\imath \, {\bf k_{m,2}} . {\bf r} } 
    \end{eqnarray*}
    
    \noindent In the latter, $\Delta {\bf r}_m = (0,0,d)$, ${\bf k_{m,1}} = k_m \; (\sin(\theta_r), 0, \cos(\theta_r))$ and ${\bf k_{m,2}} = k_m \; (\sin(\theta_r), 0,  - \cos(\theta_r))$ with $k_m = n_m \omega /c$. Below the membrane, the electric field can be written as the sum of all transmitted beams:
    
    \begin{eqnarray*}
    E_{III, p} ({\bf r}) &=& t_{12} t_{21} \sum_{n=0}^{\infty} (r_{21})^{2n}  e^{\imath \Phi_m (n+\frac{1}{2})}  
    e^{ \imath \, {\bf k_i} .({\bf r} - \Delta {\bf r}_m  - (n+\frac{1}{2}) \Delta {\bf r})} \\
    & = & a_3 ( \theta_i, n_m, d) e^{ \imath \, {\bf k_i} .{\bf r} }
    \end{eqnarray*}

    \noindent Inserting these expressions in Eq. \eqref{definition_g}, we get the expression of the electron-interaction coupling constant for electrons travelling through the membrane:
    \begin{eqnarray*}
    g_{mem} =  \frac{ \imath \, e}{2 \hbar \omega} e^{\imath k_{i,x}.x}  && \left [ \frac{1}{ \frac{\omega}{v} - k_{i,z} }  +  \frac{a_1}{ \frac{\omega}{v} - k_{r,z}}  - \frac{a_3}{ \frac{\omega}{v} - k_{i,z}} e^{\imath (k_{i,z} - \frac{\omega}{v}) d}  \right . \\
    && \left.  +  a_{m,1}\frac{e^{ \imath (k_{m,1,z} - \frac{\omega}{v}) d} - 1}{ \frac{\omega}{v} - k_{m,1,z}}  +  a_{m,2}\frac{e^{ \imath (k_{m,2,z} - \frac{\omega}{v}) d} - 1}{ \frac{\omega}{v} - k_{m,2,z} }  \right]\\
    \end{eqnarray*}
    
    \noindent These expressions reduce to ones provided in ref. \cite{konecna_electron_2020} when considering only only one reflection/transmission at the membrane interface.
    We plot in Figure S1-b the value of $g_{mem}$ as a function of the membrane thickness and incidence angle for an illumination wavelength $\lambda_{inc} = 515$ nm. The configuration of our experiment is highlighted by the white dot visible in Figure S1-b.

    \subsection{Case of an apertured membrane}

    In the case in which the electron travels through an aperture fabricated in a metallic film deposited on a dielectric membrane, the electric field above the membrane (illuminated region) $E_{I, p} ({\bf r}, t)$ takes two different expressions depending on whether the incident wave has been reflected by the metallic film or the dielectric membrane.
    Denoting $(x,y)$ the coordinates of the electron beam in the plane of the aperture edge, neglecting the metallic film thickness and following a geometrical approximation, we can identify in the illuminated region ($z<0$) for each position of the electron beam an altitude $z_t$ which separates two different cases : reflection on the metallic film ($z < z_t $) or on the dielectric membrane ($z > z_t $). For rectangular apertures illuminated from the side (see Fig. S1-C) with their edge located at $x=0$, $z_t$ is simply given by $z_t = - x/\tan \theta_i$, $x$ being the distance to the aperture.
    
    \noindent We therefore have the following expressions for the electric field along the electron trajectory above the membrane:
    \begin{eqnarray}
    E_{I, p} ({\bf r}, t) &=& E^{Au}_{I, p} ({\bf r}, t) =  e^{\imath \, {\bf k_i} . {\bf r} } + r_{Au}.e^{\imath \, {\bf k_r} . {\bf r} } \hspace{2.2cm} \mathrm{ for \;  z < z_t}  \\
    E_{I, p} ({\bf r}, t) &=& E^{mem}_{I, p} ({\bf r}, t) =  e^{\imath \, {\bf k_i} . {\bf r} } +   a_1 ( \theta_i, n_m, d) e^{\imath \, {\bf k_r} . {\bf r} } \hspace{.6cm} \mathrm{ for \; z >z_t} \\ \nonumber
    \end{eqnarray}
    Furthermore, the integration limits along the electron trajectory must take into account the shadowing effect of the metallic film in the region below the sample.
    The entry point in the shadow zone is $z_s = d + (x - d \tan \theta_t ) /\tan \theta_i$.
    This finally leads to a modified expression for $g$ :
    \begin{eqnarray*}
    g(x,y) =  \frac{\imath \, e}{2 \hbar \omega} e^{\imath k_{i,x}.x}  && \left [ \frac{1}{\frac{\omega}{v} - k_{i,z} }  +  \frac{r_{Au} }{\frac{\omega}{v} - k_{r,z} }  e^{\imath (k_{r,z} - \frac{\omega}{v}) z_t}   + \frac{a_{1}}{\frac{\omega}{v} - k_{r,z} }  ( 1 - e^{\imath (k_{r,z} - \frac{\omega}{v}) z_t}  )   \right . \\
    &&   +  a_{m,1}\frac{e^{ \imath (k_{m,1,z} - \frac{\omega}{v}) d} - 1}{ \frac{\omega}{v} - k_{m,1,z}}  +  a_{m,2}\frac{e^{ \imath (k_{m,2,z} - \frac{\omega}{v}) d} - 1}{ \frac{\omega}{v} - k_{m,2,z} }  \\
    && \left.    + a_3 \frac{e^{\imath (k_{i,z} - \frac{\omega}{v}) z_s} - e^{\imath (k_{i,z} - \frac{\omega}{v}) d}}{k_{i,z} - \frac{\omega}{v}}  \right]\\
    &=& g_{mem} +  \frac{r_{Au} - a_1}{\frac{\omega}{v} - k_{r,z} }  e^{\imath (k_{r,z} - \frac{\omega}{v}) z_t}  +  a_3 \frac{e^{\imath (k_{i,z} - \frac{\omega}{v}) z_s} }{k_{i,z} - \frac{\omega}{v}}
    \end{eqnarray*}
    
    \noindent The modulation of the electron-light coupling constant arises from the contrast in reflectivity and shadowing effect of the metallic film, respectively accounted for by the second and third terms of the previous expression
    For apertures of arbitrary shapes, $z_t$ and $z_s$ must be calculated for each position of the electron beam inside the aperture.
    We plot in Figure S1-D the electron-light coupling constant predicted by this analytical model for a metallic film edge as a function of the incidence angle and distance to the film edge.
    The value of $g$ has been normalized to the electron-light coupling constant $g_{mem}$ of the bare membrane in our experiments ($\lambda_{inc} = 515$ nm, $\theta_i = 35^{\mathrm{\circ}}$).
    Clear modulations are visible with a typical length-scale decreasing with the incidence angle.
    
    \subsection{Diffraction by the edge of a metallic film}

    \begin{center}
    \begin{figure}[htp]
    \centering
    \includegraphics[width=\columnwidth]{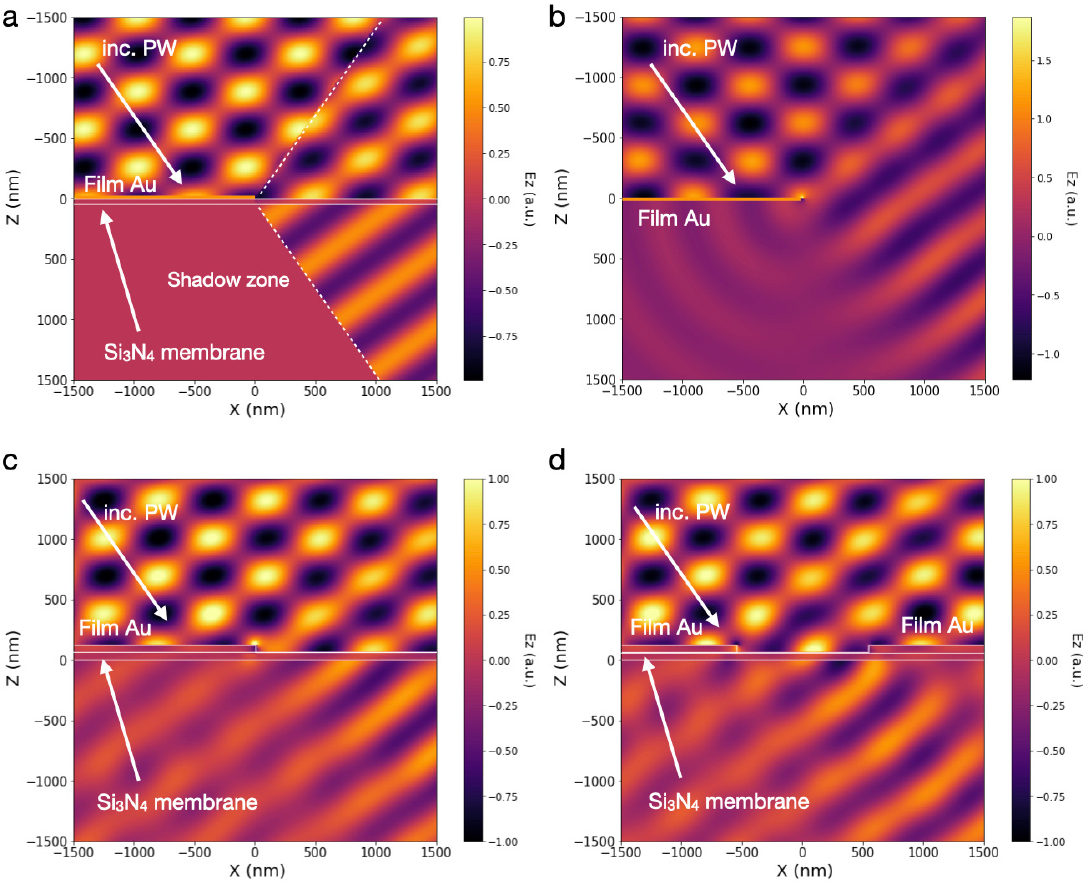}
    \caption{(Color Online) a) Electric field in the vicinity of a dielectric membrane half covered by a metallic film : analytical model.
    b) Electric field in the vicinity of a semi-infinite metallic film : Sommerfeld model.
    c) Electric field in the vicinity of a dielectric membrane half covered by a metallic film : 2D GDM numerical simulations.
    d) Electric field in the vicinity of a dielectric membrane  covered by a two parallel metallic film separated by a gap of width $w = 1100$ nm : 2D GDM numerical simulations.}
    \label{FigureS3}
    \end{figure}
    \end{center}

    The electric field considered in the analytical model is represented in Figure S2-A.
    The geometrical approximation used clearly misses the contribution from diffraction effects such as the scattering of the incident wave by the edge of the metallic film.
    Following the approach first proposed by A. Sommerfeld, it is possible to compute the electric field in the vicinity of a metallic half-plane \cite{BornWolf:1999:Book}.
    This model assumes that the metallic film is in vacuum and therefore discards the contribution from the membrane.
    The results of Figure S2-B show as expected the contribution of the scattered wave in the shadow region.

    \subsection{Electrodynamical simulations using the Green Dyadic Method}
    
    To take into account the different contributions,  we have performed 2D and 3D electrodynamical simulations using the Green Dyadic Method.
    The resolution of Mawell's equations in the framework of the GDM is based on the computation of a so-called generalized propagator which captures the entire electrodynamical response of the system.
    The computation of the generalized propagator relies on a discretization of the volume of the system.
    We have used the open source python toolkit pyGDM \cite{wiecha_pygdm_2022}.
    The latter includes tools to easily derive several physical quantities such as extinction, scattering and absorption cross-sections, far-field patterns, the electric and magnetic near-field and its multipole decomposition, the decay-rates / LDOS inside and in the vicinity of a structure, or the heat dissipated by a nanoparticle.\\
    
    \noindent Figure S2-C shows the result of a 2D GDM simulation in which we simulate the optical response of a $\mathrm{Si}_{3}\mathrm{N}_{4}$ membrane of thickness $d = 50$ nm half covered by a gold film of thickness 50 nm.
    The electric field predicted by the full electrodynamical simulations is generally in good agreement with the analytical model.
    As expected, the numerical simulations also account for the diffraction by the film edge.
    Figure S2-D displays the results of 2D-GDM simulations taking into account two metallic films separated by a gap of width $w = 1100$ nm.
    The numerical simulations displayed in this study typically involves a volume discretization with 5000-10000 mesh cells.

    \subsection{Contributions of scattering by the metallic film edge and dielectric membrane to the modulations of the electron-light coupling constant}

\begin{center}
    \begin{figure*}[htp]
    \centering
    \includegraphics[width=14cm,angle =0.]{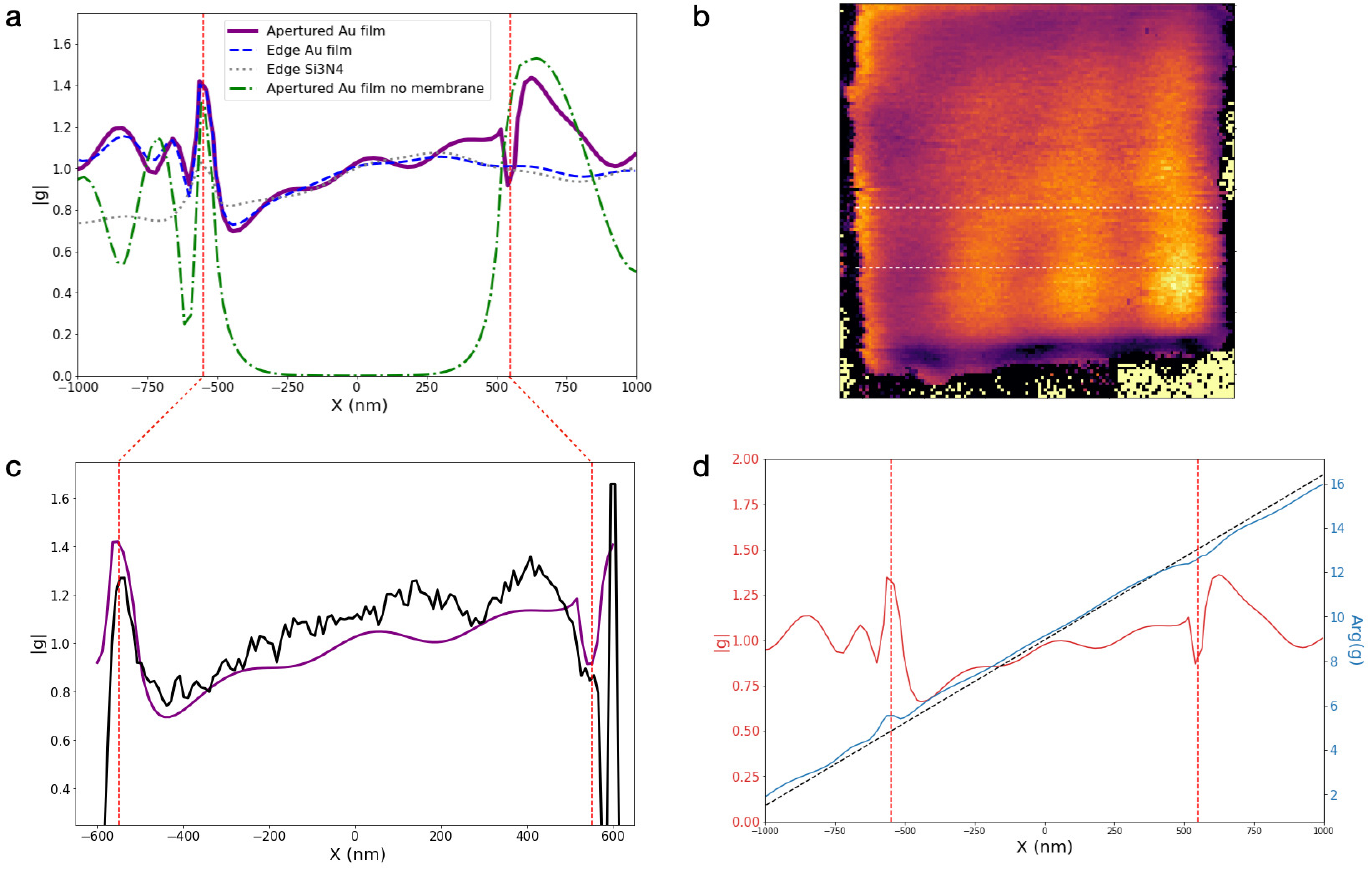}
    \caption{(Color Online) a) Caption.
    a) Profile of the electron-light coupling strength predicted by 2D-GDM simulations in the case of an apertured metalllic film without membrane in the aperture (green dash-dotted line), the edge of a $\mathrm{Si}_{3}\mathrm{N}_{4}$ (grey dotted line) or gold (blue dashed line) film half covering a silicon nitride membrane.
    b) PINEM map. The experimental profile compared to the electrodynamical simulations in c) is the average of the profiles measured between the two white lines.
    c) Comparison between the electron-light coupling strength predicted by the numerical simulations and the experimental results.
    d) Modulus and argument of the coupling strength $g$ predicted by the simulations. The black dashed line corresponds to the term $e^{\imath k_{i,x}.x }$ term of the analytical expression of $g_{mem}$ shifted by a constant factor.}
    \label{FigureS4}
    \end{figure*}
    \end{center}

    We here address the relative contributions of the scattering by the film edge and the interference effects in the dielectric membrane on the modulations of the electron-light coupling constant.
    Figure \ref{FigureS4}-a displays the result of 2D-GDM computation performed on an apertured film with no membrane, a silicon nitride membrane half covered by either a gold film or an additional silicon nitride film or a silicon nitride membrane covered by a gold film in which a gap of 1100 nm width has been fabricated.
    The values have been normalized to the coupling constant of a bare $\mathrm{Si}_{3}\mathrm{N}_{4}$ membrane.
    in the absence of membrane, no electron-light coupling is possible when the electron probe is far enough from the aperture edge. This was confirmed experimentally in Figure S1-a.
    The presence of an edge yields an additional scattered wave that cause spatial modulations of the electron-light coupling strength.
    Finally, we notice that the variation of the coupling strength in the case of the aperture varies similarly as the gold film edge until the electron beam gets closer to the other edge where the influence of the latter becomes noticeable.
 
    \noindent In figure \ref{FigureS4}-c we plot on the same graph the electron-light coupling strength computed using the 2D-GDM simulations and the experimental results confirming the good agreement.
    In figure \ref{FigureS4}-d, the modulus and argument of the electron-light coupling strength are displayed together with the phase spatial variations that would correspond to the $e^{\imath k_{i,x}.x}$ term of Equation (\ref{def_gmem_main}) of $g_{mem}$ of the main text.
    This shows that inside the apertured metallic film, the phase of the electron-light coupling constant $g$ evolves similarly as a bare membrane.
    The influence of the aperture edges is mostly limited to a variation of $\pm 20 \%$ of the modulus of the coupling.

\end{document}